\documentstyle[psfig]{mn}

\newif\ifAMStwofonts

\ifoldfss
  \ifCUPmtlplainloaded \else
    \NewTextAlphabet{textbfit} {cmbxti10} {}
    \NewTextAlphabet{textbfss} {cmssbx10} {}
    \NewMathAlphabet{mathbfit} {cmbxti10} {} 
    \NewMathAlphabet{mathbfss} {cmssbx10} {} 
  \fi
  \ifAMStwofonts
    \ifCUPmtlplainloaded \else
      \NewSymbolFont{upmath} {eurm10}
      \NewSymbolFont{AMSa} {msam10}
      \NewMathSymbol{\upi}     {0}{upmath}{19}
      \NewMathSymbol{\umu}     {0}{upmath}{16}
      \NewMathSymbol{\upartial}{0}{upmath}{40}
      \NewMathSymbol{\leqslant}{3}{AMSa}{36}
      \NewMathSymbol{\geqslant}{3}{AMSa}{3E}

    \fi
  \fi
\fi 

\ifnfssone
  \newmathalphabet{\mathit}
  \addtoversion{normal}{\mathit}{cmr}{m}{it}
  \addtoversion{bold}{\mathit}{cmr}{bx}{it}
  \newmathalphabet{\mathbfit} 
  \addtoversion{normal}{\mathbfit}{cmr}{bx}{it}
  \addtoversion{bold}{\mathbfit}{cmr}{bx}{it}
  \newmathalphabet{\mathbfss} 
  \addtoversion{normal}{\mathbfss}{cmss}{bx}{n}
  \addtoversion{bold}{\mathbfss}{cmss}{bx}{n}
  \ifAMStwofonts
    \ifCUPmtlplainloaded \else
      \UseAMStwoboldmath
      \makeatletter
      \new@mathgroup\upmath@group
      \define@mathgroup\mv@normal\upmath@group{eur}{m}{n}
      \define@mathgroup\mv@bold\upmath@group{eur}{b}{n}
      \edef\UPM{\hexnumber\upmath@group}
      \new@mathgroup\amsa@group
      \define@mathgroup\mv@normal\amsa@group{msa}{m}{n}
      \define@mathgroup\mv@bold\amsa@group{msa}{m}{n}
      \edef\AMSa{\hexnumber\amsa@group}
      \makeatother
      \mathchardef\upi="0\UPM19
      \mathchardef\umu="0\UPM16
      \mathchardef\upartial="0\UPM40
      \mathchardef\leqslant="3\AMSa36
      \mathchardef\geqslant="3\AMSa3E
    \fi
  \fi
\fi 

\ifnfsstwo
  \DeclareMathAlphabet{\mathbfit}{OT1}{cmr}{bx}{it}
  \SetMathAlphabet\mathbfit{bold}{OT1}{cmr}{bx}{it}
  \DeclareMathAlphabet{\mathbfss}{OT1}{cmss}{bx}{n}
  \SetMathAlphabet\mathbfss{bold}{OT1}{cmss}{bx}{n}
  \ifAMStwofonts
    \ifCUPmtlplainloaded \else
      \DeclareSymbolFont{UPM}{U}{eur}{m}{n}
      \SetSymbolFont{UPM}{bold}{U}{eur}{b}{n}
      \DeclareSymbolFont{AMSa}{U}{msa}{m}{n}
      \DeclareMathSymbol{\upi}{0}{UPM}{"19}
      \DeclareMathSymbol{\umu}{0}{UPM}{"16}
      \DeclareMathSymbol{\upartial}{0}{UPM}{"40}
      \DeclareMathSymbol{\leqslant}{3}{AMSa}{"36}
      \DeclareMathSymbol{\geqslant}{3}{AMSa}{"3E}
    \fi
  \fi
\fi 

\ifCUPmtlplainloaded \else
  \ifAMStwofonts \else 
    \def\upi{\pi}
    \def\umu{\mu}
    \def\upartial{\partial}
  \fi
\fi

\title[Gravitational Potential Perturbations]
	{Importance of Perturbed Gravitational Potentials in Differentially
	Rotating Newtonian Stars}

\author[S. Karino]
       {S. Karino \\
	Department of Earth Science and Astronomy,\\
        Graduate School of Arts and Sciences,\\
        University of Tokyo, Komaba, Meguro, Tokyo 153-8902, Japan}
\date{Accepted ???? Month ??.
      Received ???? Month ??;
      in original form ???? Month ??}

\pagerange{\pageref{firstpage}--\pageref{lastpage}}
\pubyear{????}

\begin{document}

\maketitle

\label{firstpage}

\begin{abstract}
%
It is usually believed that the Cowling approximation 
can give satisfactory solutions if the stars 
have soft equations of state
and/or
if the strongly general relativistic stars are treated in the case of
rigid rotation.
Since, however, there have been
no systematic studies about the accuracy of the Cowling approximation
for differentially rotating compressible stars, 
we investigate 
eigenfrequencies and eigenfunctions of the 
oscillation modes
in {\it rapidly} and {\it differentially} rotating compressible stars
by employing the exact method including full-perturbations
and the Cowling approximation. 

We have found that the Cowling approximation for f-mode oscillations
is not a good approximation
in rapidly and differentially rotating stars,
although 
rapid rotation makes this 
approximation better for rigidly rotating stars.
This result suggests that we must be careful when we apply the Cowling
approximation to differentially rotating stars even in the framework of
general relativity.
On the other hand, the approximation will work well
for r-modes even if the star is rotating differentially.
Therefore, the Cowling approximation can be used as a strong tool for the
investigation of r-mode oscillations in the general relativistic framework
that it is difficult to compute including the perturbations of gravity.

\end{abstract}


\section{Introduction}

Concerning the oscillations of spherical stars, Cowling (1941) 
started to use an approximation that 
neglect the perturbed gravitational potential 
to rewrite the 
system of equations into
easier form.
By using this approximation (later it is called the Cowling 
approximation), he found that there are three types of oscillations; the high 
frequency acoustic mode (p-mode), the low frequency gravity mode (g-mode), 
and the intermediate fundamental mode (f-mode).   

In the last two decades, the Cowling approximation has been applied to general
relativistic compact stars.
In spherical stars, McDermott, Van Horn, and Scholl (1983) proposed
the scheme in which 
{\it all} components of the metric perturbations are neglected.
On the other hand, Finn (1988)
proposed the formulation which neglects
{\it most} components of the 
metric perturbations except for the $(r,t)$ component.
The accuracies of 
these two approaches were compared by Lindblom and Splinter (1990)
for mainly p-modes, and 
it was found that generally the approximation neglecting {\it all} components 
of the metric perturbations is better than the other one, while there is a 
possibility that the latter becomes better when the g-modes are considered.
Hereafter, ``the Cowling approximation'' means the one that neglects {\it all}
components of the metric perturbations.

It is also shown that the Cowling approximation is reliable in slowly rotating 
general relativistic stars (Kojima 1997). Furthermore, Ipser and 
Lindblom (1992) proposed the 
formulation of applying the Cowling approximation to 
rapidly rotating relativistic stars.  The accuracy of the Cowling 
approximation in 
rotating stars was studied by Yoshida and Kojima (1997).
They compared the two sets of solutions obtained by using the 
procedure produced by Ipser and Lindblom (1992) and by solving the exact
equations for the f- and p-mode oscillations of nearly spherical stars 
with slow rotation.
According to their results, solutions of the Cowling 
approximation agree 
well with the exact solutions for
slowly rotating and highly relativistic models.
Recently, Yoshida and coworkers have used the Cowling approximation
to find neutral points of the f-mode oscillations
of rapidly rotating polytropes (Yoshida and Eriguchi 1997, 1999, 2001; 
Yoshida, Rezzolla, Karino, and Eriguchi 2002).
Also they checked the accuracy of the approximation in relativistic stars
and found the conditions that the approximation becomes accurate
(Yoshida and Eriguchi 1997, 1999).

Our knowledges about the Cowling approximation to the oscillations
of rotating stars thus far can be summarized as follows:\\
The Cowling approximation becomes more accurate \\
%
%
(1)
when the compressibility of the star is sufficiently high,\\
%
%
(2)
when the gravity of the star is sufficiently strong.\\
%
It may be natural that the Cowling approximation becomes more accurate if 
the effect of the gravity does not work effectively to the fluid mass.
For the simple oscillations such as f-mode, the amplitude of the mode
becomes a simple monotonically increasing function of a stellar radius,
and takes its maximum at the surface.
Here, when we consider the density distribution of
a compressible star, the density of surface region becomes relatively low.
In other words, there are only small masses in the region where the mode
amplitude gets large, in the centrally condensed star.
Also, effectively, the strong gravitation in general relativity (GR)
can have the same effect.
Consequently, the effects of compressibility and strong gravity make the
accuracy of the Cowling approximation better.\\
and\\
(3)when the rotation of the star is sufficiently rapid,
when the star rotates rigidly.\\
Here, notice that the effect of differential rotation on the approximation
has not studied, therefore we will focus on it.

From the astrophysical points of view,
we should take additional effects 
into account. For example, some stars are rotating differentially. In 
particular, newly born neutron stars may rotate differentially (see e.g. 
Liu and Lindblom 2001, Shibata and Uryu 2000,
Dimmelmeier, Font and M\"{u}ller 2002). Young neutron stars have been 
considered to be  astrophysically interesting objects because they can be 
candidates of sources of gravitational waves due to the rotational instability 
(Chandrasekhar 1970, Friedman and Schutz 1978ab).
However, if stars rotate 
differentially, the eigenfunctions of oscillations as well as the equilibrium 
configurations will be deformed considerably (see e.g. Eriguchi and 
M\"{u}ller (1985) for Newtonian equilibrium stars; Komatsu, Eriguchi, and 
Hachisu (1988) for relativistic equilibrium stars; Karino, Yoshida, and 
Eriguchi (2001) for eigenfunctions of oscillations of Newtonian stars.)
If one would like to apply the Cowling approximation to such differentially
rotating stars, it is necessary to estimate the applicability of the Cowling 
approximation to them. 
In this paper we will study the accuracy of the Cowling approximation
in rotating Newtonian stars by comparing the results obtained by the
Newtonian Cowling approximation with those 
by solving the exact equations for the perturbed equations. 

The plan of this paper is as follows.
In the next section, we will briefly summarize the scheme to solve the 
linearized perturbed equations for differentially rotating stars with and 
without the gravitational perturbation. Our numerical results will be 
presented in Section 3. 
In Section 4, the accuracy of the Cowling approximation will be discussed.
In the last section, we will summarize our present study and make a brief 
comment on the applicability of the Cowling approximation to GR.
Throughout this paper, the ordinary spherical polar coordinate 
system $(r, \theta, \varphi)$ is used.

\section{Method}



The motivation of this paper is to estimate the accuracy of the Cowling
approximation. For this purpose, we will solve the linearized equations of
perturbed fluid in Newtonian gravity and numerically obtain the eigenvalues
and eigenfunctions of oscillation modes. Then we will compare them with the 
results computed by neglecting the gravity perturbation, $\delta \phi$, from
the basic equations (the Cowling approximation).
The method of solving the full system of the basic equations for 
the oscillations of rotating stars is shown in
Karino, Yoshida, Yoshida and Eriguchi (2000).
Hence, in this section, we will briefly explain the solving method.

The equilibrium states of axisymmetric 
polytropes can be obtained by the method proposed by Eriguchi
and M\"{u}ller (1985).  The polytropic relation is expressed as
\begin{equation}
p = K \rho^{1+\frac{1}{N}},
\end{equation}
where $p$, $\rho$, $K$ and $N$ are the pressure, the density, the polytropic 
constant and the polytropic index, respectively. In this paper, we compute 
polytropes with $N=$ 0.5, 1.0, and 1.5. 
Such relatively stiff polytrope can be used as the models of neutron stars.

We will use the following rotation law (j-constant rotation 
law):
\begin{equation}
\Omega = \frac{\Omega_0 A^2}{(r \sin \theta)^2 + A^2},
\end{equation}
where $\Omega$ and $\Omega_0$ are the angular velocity and the angular 
velocity on the rotation axis. The parameter $A$ represents a degree of 
differential rotation. A small value of $A$ corresponds to a strongly 
differential rotation, while in the limit of $A \to \infty$ the rotation law 
becomes that of rigid rotation. 

The eigenfrequencies and eigenfunctions of the modes can be obtained by
solving the perturbed fluid equations.
Here, we assume small perturbations and linearize the system of equations.
The linearized quantities are expanded as follows:
\begin{equation}
\delta f (r,\theta,\varphi,t)
= \sum_m \exp(i(\sigma t - m \varphi)) f_m(r,\theta),
\end{equation}
where $f$ and $\delta f$ are a certain physical quantity and its Eulerian 
perturbation, respectively.
In this expansion, $\sigma$ denotes the eigenvalue of the oscillation mode
and $m$ is the azimuthal wave number.
Here, we only consider the $m=2$ f- and r-mode for two reasons:
(1) from the astrophysical point of views, the $m=2$ mode is the most 
interesting because it has the fastest growth rate in rapidly rotating
stars, and (2) it has been said 
that the accuracy of the Cowling approximation becomes the worst for
the $m=2$ mode.
Actually, the eigenfrequencies of radial and quasi-radial-modes have been
studied by nonlinear computations and it has been shown that the frequency
obtained by the Cowling approximation is considerably different from the one
given by full GR calculations even in the case with rapid rotations
(Font, Goodale, Iyer, Miller, Rezzolla, Seidel, Stergioulas, Suen
and Tobias 2002).

On the other hand, in the Cowling approximation, the Eulerian perturbation
of the gravitational potential is dropped from the full system and solved
the equations with the remaining terms.

\section{Numerical Results}

\begin{figure}
      \psfig{file=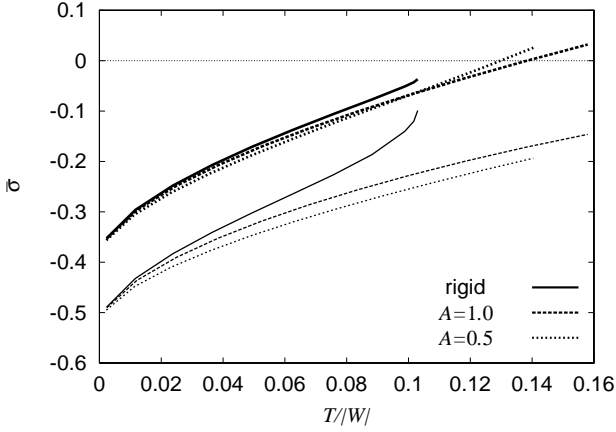,width=8.5cm,angle=0}
\caption{The eigenvalues of the $m=2$ f-mode oscillations for $N=1.0$ 
polytropes. The eigenvalue $\sigma$ is normalized as $\bar{\sigma} = \sigma 
/ \sqrt{4 \pi G \rho_c}$. Three curves of different types (solid, dashed 
and dotted) denote sequences with rigid rotation, weakly differential
rotation ($A=1.0$), and strongly differential rotation ($A=0.5$),
respectively.
The upper sequence of the curves of the same type (bold) corresponds to the 
solutions of the full perturbed equations and the lower sequence corresponds
to those of the Newtonian Cowling approximation.}
\label{fig:1}
\end{figure}

The eigenvalues of the $m=2$ f-modes for the $N=1.0$ polytrope are plotted
in Fig.~\ref{fig:1} as a function of $T/|W|$, where $T$ is the rotational 
kinetic energy and $W$ is the gravitational potential energy of the star.
Three curves of the different types correspond to the different values of the 
parameter $A$; $ A = 0.5, 1.0$, and $\infty$ (= rigid rotation).
For each 
$A$, two curves are plotted: the upper curves correspond to 
the results of the full computations and the lower 
to the approximated solutions.
In the case of $A = 1.0$ and 1.5 including the perturbation of gravity,
when the rotation of the star get rapid, the so-called neutral point
where $\sigma = 0$ appears.
This neutral point is corresponding to the critical point of secular
instability due to gravitational radiation reaction
(Chandrasekhar 1970, Friedman \& Schutz 1978ab).
On the other hand, the curves given by the Cowling approximation have
smaller, finite values.
This means that we can not estimate the critical points of
secular instability when we apply the Cowling approximation.

\begin{figure}
       \psfig{file=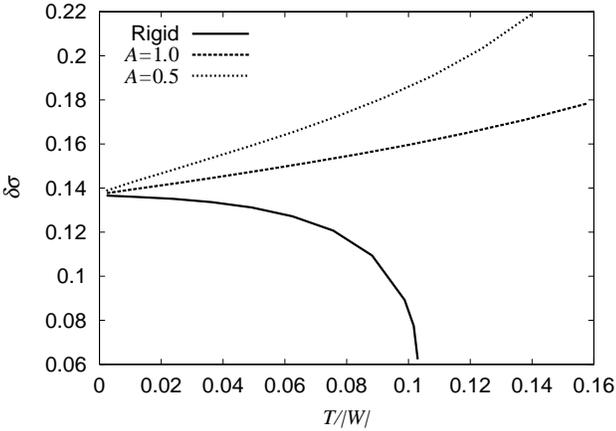,width=8.5cm,angle=0}
\caption{The differences of the eigenfrequencies between the results of the
full perturbation and those of the Newtonian Cowling Approximation
($\delta \sigma = |\sigma_{\rm{full-perturb.}} - \sigma_{\rm{approx.}}|$)
for rigid rotation (solid), $1.0$
(dashed), and $A=0.5$ (dotted). The terminal point of the rigid rotation
sequence corresponds to the mass shedding limit.}
\label{fig:2}
\end{figure}

The differences of the eigenfrequencies between the results of the full 
perturbed systems and the 
Cowling approximation are shown in 
Fig.~\ref{fig:2}. From this figure, it is clear that  
the accuracy of the Cowling approximation becomes better as the stellar
rotation gets faster for rigidly rotating polytropes.
For differentially rotating
polytropes, however, the accuracy of the approximation becomes worse
and worse as the stellar rotation becomes faster.

The comparison between the approximated results and full perturbed results
have been done for rigidly rotating polytrope by Jones, Andersson and
Stergioulas (2002).
They have carried out time evolving computations of the linearized perturbed
equations of rapidly and rigidly rotating Newtonian polytrope with the
Cowling approximation, and compared their results about f- and r-modes with
the previous results.
They have checked the accuracy of the Cowling approximation
and confirmed certain amount of errors of eigenfrequencies between them.
Their results for rapidly rotating stars (Figure 4 in Jones et al. 2002)
is well-matched with our result for rigid rotation case (Figure \ref{fig:1}),
although they have computed upper branch of f-mode and our results are for
lower branch.
This similarity shows our result is consistent.

\begin{figure}
       \centering\leavevmode
	\psfig{file=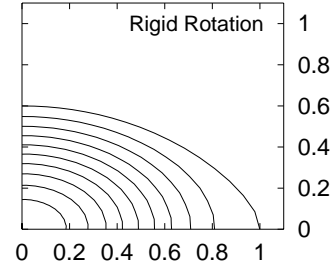,width=7.0cm,angle=0}	
\caption{Density contours of a rigidly rotating star with the polytropic 
index $N=1.0$ and $r_{\rm p}/r_{\rm e} = 0.6$ in the meridional plane. 
The $i$th contour from the 
center denotes the isodensity curve with 
$\rho = (1 - 0.1 \times i) \rho_{\rm{max}}$. Here $\rho_{\rm max}$ is the
maximum density.}
\label{fig:3}
\end{figure}

\begin{figure}
       \centering\leavevmode
        \psfig{file=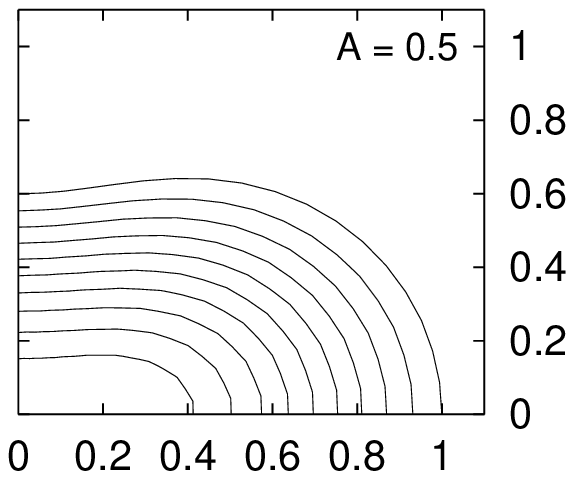,width=7.0cm,angle=0}	
\caption{Same as Fig.~\ref{fig:3} but for differential rotation ($A=0.5$).}
\label{fig:4}
\end{figure}

There are two reasons why the accuracy of the Cowling approximation becomes 
worse for differential rotation. The first reason is related to the
density distribution of the equilibrium stars. In Figs.~\ref{fig:3} and 
\ref{fig:4}, equi-density contours of the rigidly and differentially rotating 
stars with $N = 1.0$ and $r_{\rm p}/r_{\rm e} = 0.6$ are shown, respectively. 
Here $r_{\rm p}$ and $r_{\rm e}$ are the polar radius and the equatorial
radius, respectively. 
By comparing Fig.~\ref{fig:3} and Fig.~\ref{fig:4}, 
in the rapidly and 
differentially rotating star, the dense region near the center {\it extends} 
to the outer region. Remember that the amplitude of the f-mode oscillation is 
significant in the outer region of the star.
Then, in the differentially rotating star, the gravitational potential
perturbation comes to play a relatively important role, because the region
with high-amplitude is dense.
Therefore, the Cowling approximation which 
neglects the perturbed gravitational potential from the basic equations gives
less accurate solutions.

\begin{figure}
       \centering\leavevmode
       \psfig{file=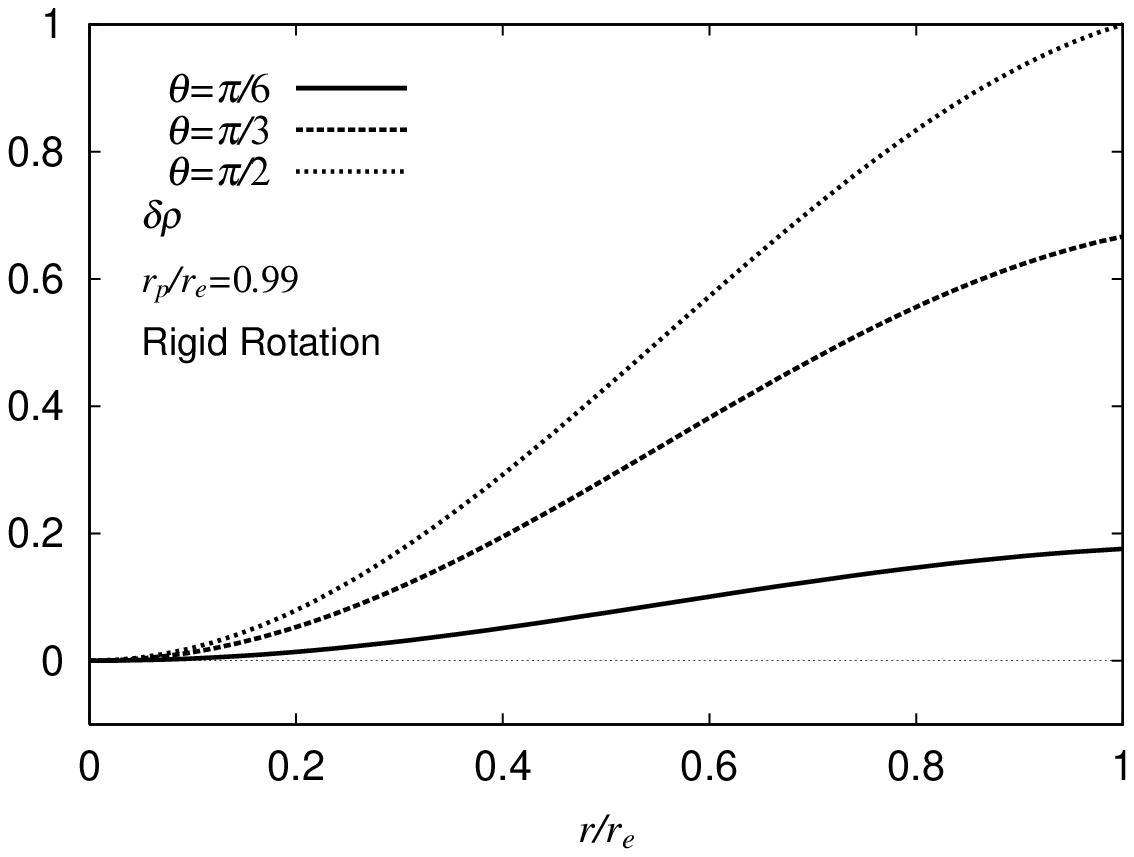,width=6.5cm,angle=0}
       \psfig{file=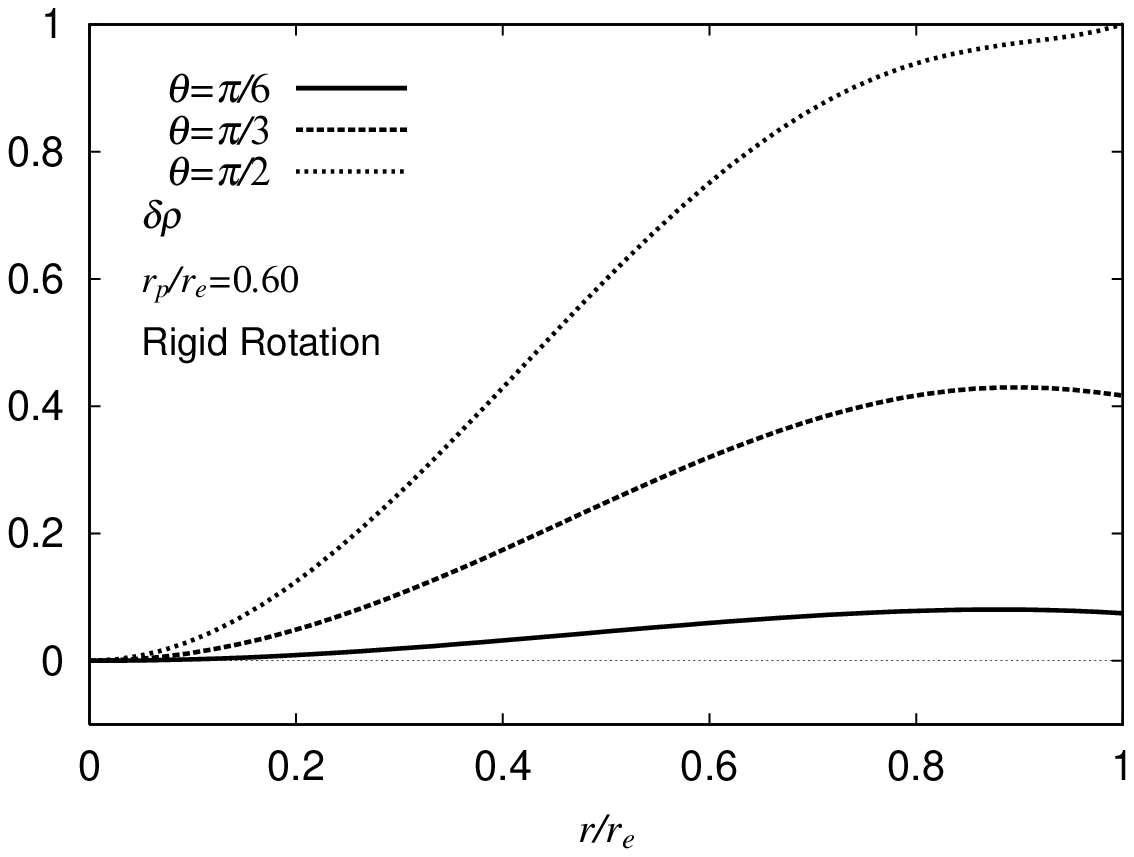,width=6.5cm,angle=0}	
\caption{The eigenfunctions of the density perturbation for the $m=2$ f-mode
oscillations in the rigidly rotating polytropes with $N = 1.0$ and
$r_{\rm p}/r_{\rm e} = 0.99$ (upper) and $r_{\rm p}/r_{\rm e} = 0.60$ 
(lower). The horizontal axis denotes the relative distance from the rotation
axis, $r /r_{\rm s}(\theta)$, and the vertical axis is the normalized 
amplitude of the perturbed density. Here $r_{\rm s}(\theta)$ denotes the 
surface of the unperturbed equilibrium configuration. Three different curves 
correspond to the eigenfunctions on the spokes of $\theta = \pi/6, \pi/3$ 
and $\pi/2$ directions.
}
\label{fig:5}
\end{figure}

\begin{figure}
       \centering\leavevmode
       \psfig{file=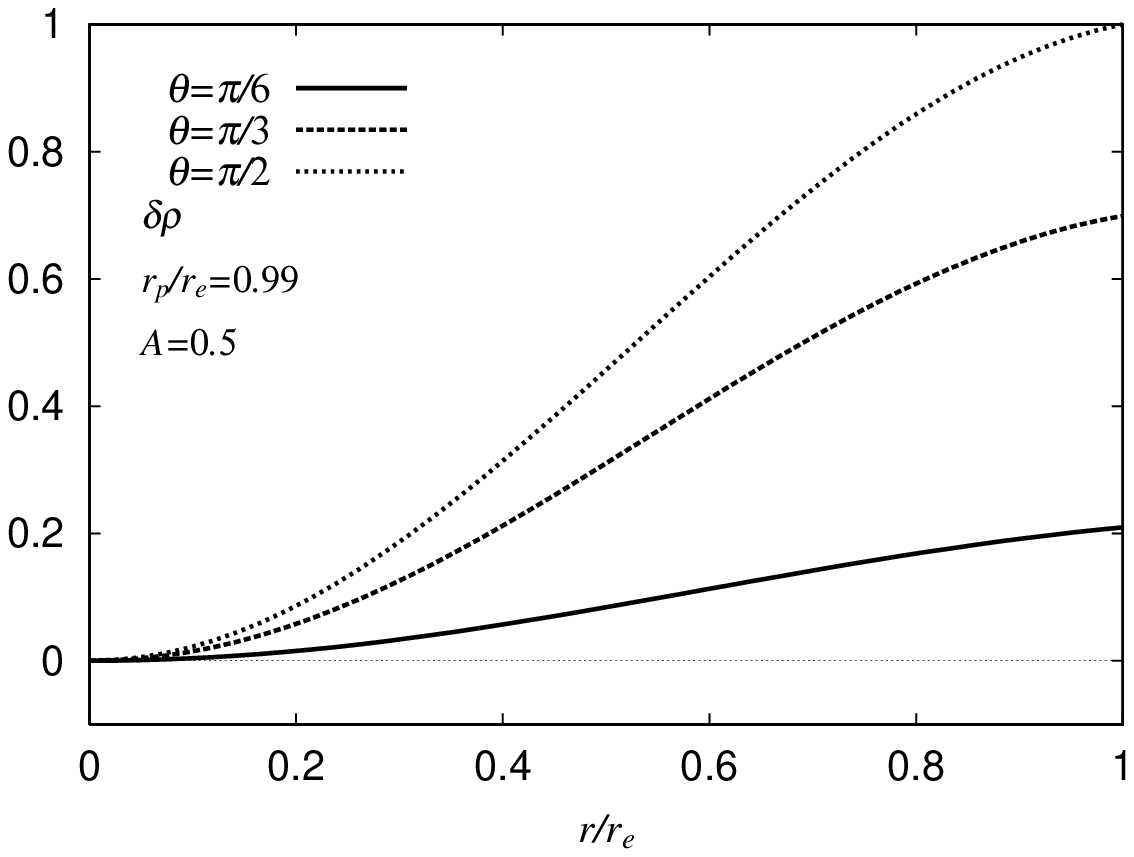,width=6.5cm,angle=0}
       \psfig{file=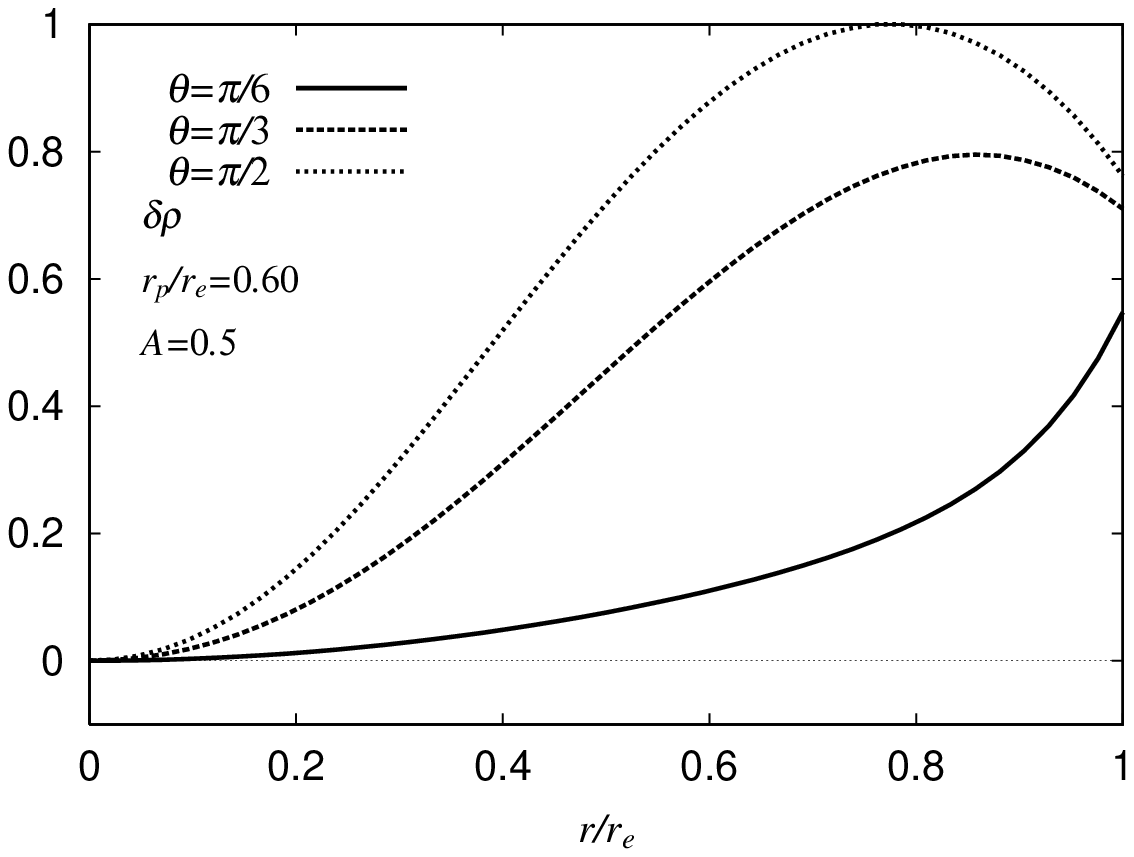,width=6.5cm,angle=0}	
\caption{Same as Fig.~\ref{fig:5} but for differential rotation ($A = 0.5$).
}
\label{fig:6}
\end{figure}

The second factor is the behavior of the eigenfunction due to differential
rotation. In Figs.~\ref{fig:5} and \ref{fig:6}, the density perturbation 
$\delta \rho$ is shown as a function of the relative distance from the
rotation axis for rigidly (Fig.~\ref{fig:5}) and differentially 
(Fig.~\ref{fig:6}) rotating $N=1.0$ polytropes. 
In the upper panels of these figures denote the eigenfunctions
for the almost spherical configurations whose axis
ratio is $r_{\rm p}/r_{\rm e} = 0.99$, while the lower panels show
those for the rapidly rotating stars with $r_{\rm p}/r_{\rm e} = 0.60$.
The eigenfunctions in those figures are normalized by the
maximum amplitude of the density perturbation in the star. 
From Fig.~\ref{fig:5}, the eigenfunctions of rigidly rotating polytropes are 
the monotonically increasing functions of the distance from the rotation axis 
and their amplitudes become the maximum at the surface.
This behaviors are common with the previous results given in the cases of
spherical stars and/or rigidly rotating stars.
In Fig.~\ref{fig:6}, however, the behavior of the 
eigenfunctions drastically change. When the stellar rotation is slow, the 
qualitative behaviors of the eigenfunctions
that increase monotonically with the stellar radii
are the same as those
of rigidly rotating polytropes. When the stellar rotation is sufficiently 
rapid, the maximum values of the eigenfunctions appear in the inner region, 
not at the surface. This result means that in rapidly and differentially
rotating stars the amplitudes of the modes in the high density region 
are rather large and that the effects of the gravitational perturbation
become important. 

\begin{figure}
       \centering\leavevmode
       \psfig{file=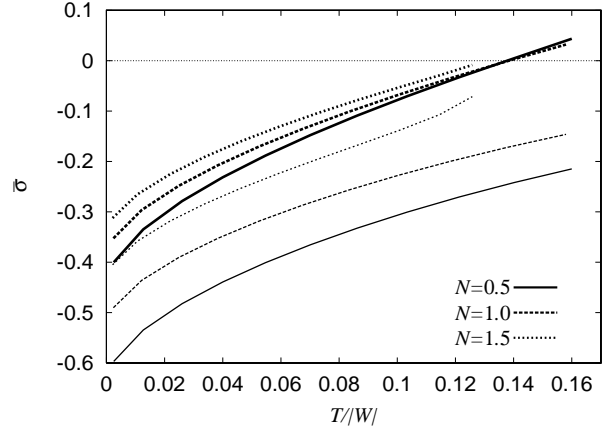,width=8.5cm,angle=0}
\caption{Same as Fig.\ref{fig:1} but for sequences with $A = 1.0$.
Three curves of different types (solid, dashed and dotted) denote sequences 
with $N = 0.5, 1.0$, and 1.5, respectively. }
\label{fig:7}
\end{figure}

In order to see the dependence of the accuracy of the Cowling approximation
on the equations of state in differentially rotating stars, the eigenvalues 
of the $m=2$ f-modes in differentially rotating ($A = 1.0$) stars, with 
several polytropic indices are also computed. The eigenvalues as functions 
of $T/|W|$ are shown in Fig.~\ref{fig:7}. Three curves of different types 
correspond to different models with $N=0.5$, $N=1.0$, and $N=1.5$. 
Similarly as shown in Fig.~\ref{fig:1}, the upper sequences of the pair
curves show the eigenvalues for the full perturbed systems, while the 
lower sequences correspond to the results of the Cowling approximation. 

\begin{figure}
       \centering\leavevmode
       \psfig{file=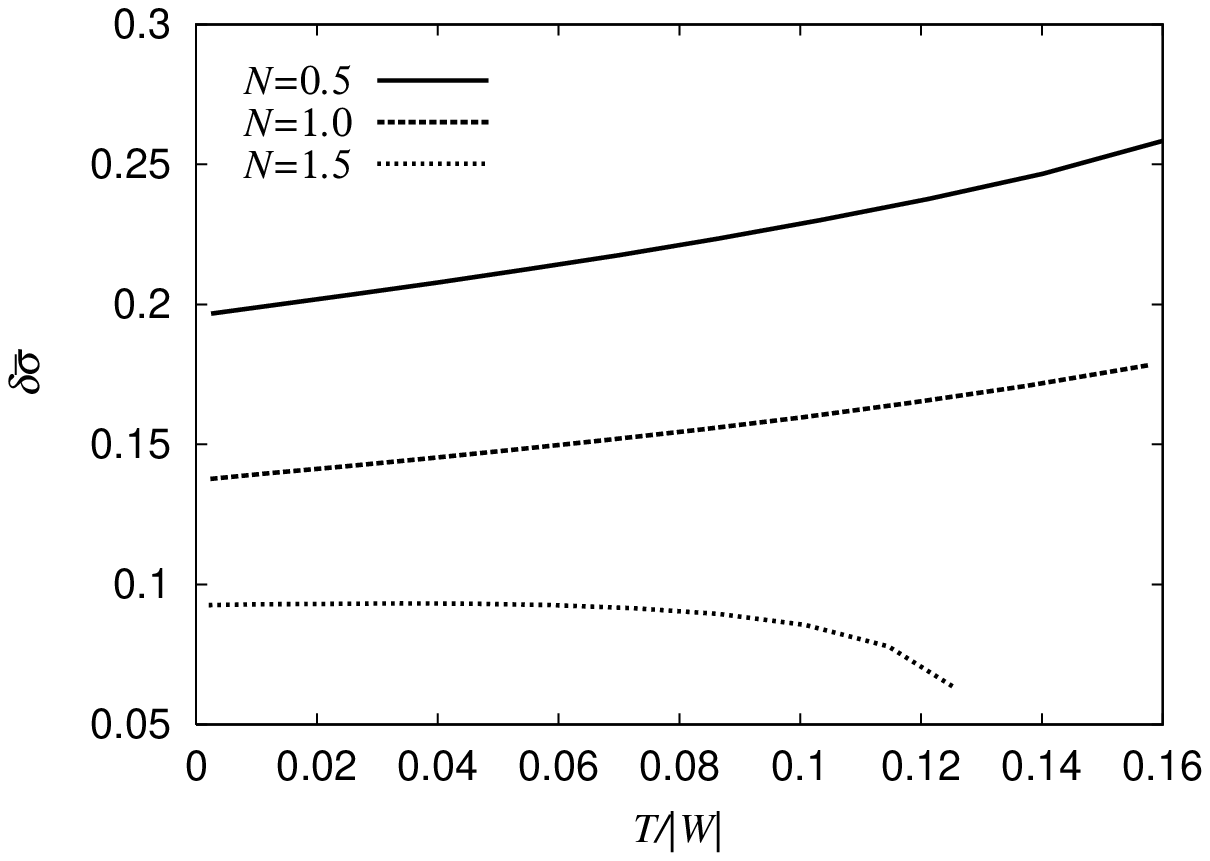,width=8.5cm,angle=0}
\caption{Same as Fig.\ref{fig:2} but for sequences with $A = 1.0$.
Three curves of different types represent sequences with  $N=0.5$ (solid), 
$N = 1.0$ (dashed), and $N=1.5$ (dotted). }
\label{fig:8}
\end{figure}

In Fig.~\ref{fig:8}, the differences of the eigenfrequencies between
the solutions of the full systems and those of the Cowling approximation
are shown as in Fig.~\ref{fig:2}.
For rigidly rotating stars, the high compressibility makes 
the accuracy of the approximation better, because the density
distributions in highly compressible star is considerably condensed
to the central region, so the density will be significantly low in
the outer region where the amplitudes of the oscillations become large.
In Fig.~\ref{fig:7} and \ref{fig:8}, almost the same tendency with the
previous one can be observed.
For $N=1.5$ polytrope, the matter of the star is more centrally condensed
compared with the smaller values of $N$.
Hence, even if the star is rotating differentially with certain degree, 
the accuracy of the Cowling approximation 
does not become inaccurate. 
On the other hand, for $N=0.5$ model, since the density of the star does
not drop significantly in the outer region, the accuracy of the Cowling
approximation becomes considerably inaccurate  mainly due to the effect of 
differential rotation.

\section{Discussion}
\subsection{f-mode}

In the previous section, we have compared the results given by  
the system of the full perturbed equations with that in
the Cowling approximations, and we have found that the accuracy of
the approximation becomes worse 
when we consider rapidly and differentially rotating stars.
And the degree of declination of the accuracy depends on the strength
of differential rotation. Two reasons for this tendency can be considered. 
The first reason will be the prolongational effect of the equilibrium density 
distribution along equatorial plane, by the presence of differential rotation. 
The second one will be the 
change that the maximum points of the eigenfunctions appear 
in the interior regions of the star, not surface regions.

\begin{figure}
       \centering\leavevmode
       \psfig{file=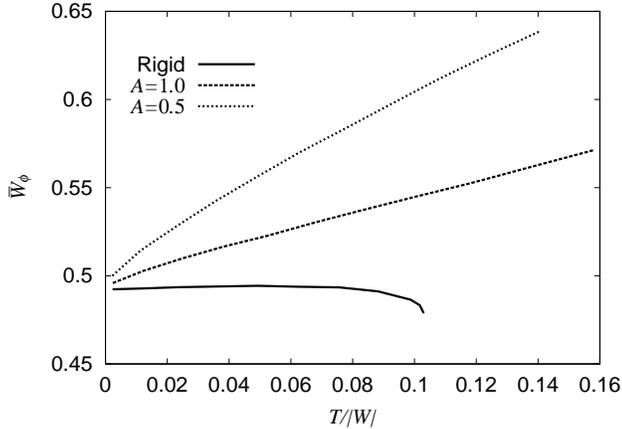,width=8.5cm,angle=0}
\caption{The values of $W_{\phi}$ for $N=1.0$ polytropes.
Results for rigid rotation, weakly differential rotation ($A = 1.0$),
and strongly differential rotation ($A = 0.5$) are shown by solid line, 
dashed line, and dotted line, respectively.
}
\label{fig:9}
\end{figure}

In this section, we will {\it quantitatively} 
estimate these effects of differential rotation.
For this purpose, we will compute the gravity perturbations numerically
in Newtonian gravity.
As mentioned above, since the reliability of the Cowling approximation
becomes better when the density of the fluid components is sufficiently
low in the region where the amplitudes of the oscillation get relatively
larger, 
in order to estimate the relation between the density in an equilibrium
and the amplitudes of the potential perturbations
, we define a new quantity as follows:
\begin{equation}
W_{\phi} \equiv \int \rho \delta \bar{\phi}_m
\exp(i(\sigma t - m \varphi)) \, dV ,
\end{equation}
where, $\delta \bar{\phi} = \delta \phi / \delta \phi_{\rm{max}}$
is the perturbed gravitational potential normalized by the 
maximum amplitude of $\delta \phi$ in the star.
\footnote{
Additionally, here, since it is assumed that the equilibrium configurations
are axisymmetric and the perturbations have the dependency as 
$\exp(i(\sigma t - m \varphi))$ in the azimuthal direction, the integral
in the $\varphi$-direction,
$\int \exp(i(\sigma t - m \varphi)) \, d \varphi$,
%
%
need not to be carried out in the present context. In this meaning, we
express this quantity as $\bar{W}_{\phi}$.
}
These quantities, 
$W_{\phi}$, for
several sequences of $N = 1.0$ polytropes
are shown in Fig.~\ref{fig:9}, for $A=1.0$, and 0.5.
The horizontal axis denotes $T/|W|$. Here, actually plotted values,
$\bar{W}_{\phi}$, are normalized by the masses of the equilibrium models:
\begin{equation}
M_0 = \int \rho \, dV.
\end{equation}
%

Introducing such a quantity and comparing Figs.~\ref{fig:9}
with Fig.~\ref{fig:2}, we can see that the behavior of 
$W_{\phi}$ becomes consistent with the absolute values of $\delta \sigma$
in terms of the function of $A$ and $T/|W|$. 
These tendencies of $W_{\phi}$ strongly suggest that our conjecture
described in previous sections are correct; i.e. the accuracy of the
Cowling approximation is related with the values of $W_{\phi}$.

\subsection{r-mode}

In these days, from astrophysical point of view, r-mode oscillations in
neutron stars are noticed.
A lot of studies which suggest that r-modes can be unstable in every
rotating star and the accompanied gravitational waves can be detected
by improved interferometer as like LIGO II has been done (Andersson 1998,
Friedman and Morsink 1998, for recent works, see the review of Andersson 2002
and its references).

\begin{figure}
       \centering\leavevmode
       \psfig{file=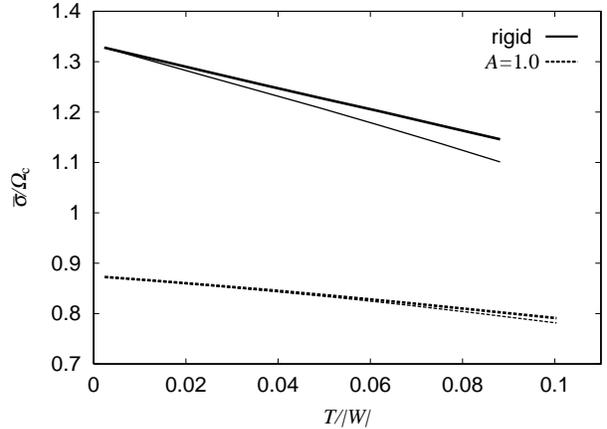,width=8.5cm,angle=0}
\caption{The same as Fig.~\ref{fig:1}. But in this case, the eigenvalues are
divided by the central angular velocity $\Omega_{\rm{c}}$ in order to
compare the results with Karino et al. (2001). The lower curves in each pair
denote the results obtained by the approximation.
}
\label{fig:11}
\end{figure}

Since the eigenfunctions of the r-mode oscillations are surface-standing,
they may not suffer from the effect of extension of the central dense regions.
Additionally the gravity perturbation is much smaller than the fluid
perturbations for r-modes (about the feature of r-mode oscillations, Saio 1982,
Yoshida and Lee 2002, for the eigenvalue of r-modes in differentially
rotating stars, see Karino et al. 2001).
Because of these reasons, the Cowling approximation 
may be accurate in the calculations of r-mode oscillations.
In order to prove this, we have carried out the same computation with f-modes
for the $m=2$ r-mode.
The results are shown in Figure \ref{fig:11}. The plot is in the same manner
with Figure \ref{fig:1}.
From this result, it can be recognized that the approximation will
give good results 
for r-mode oscillations.
This conclusion is also comparable with the assertion of Jones et al. (2002).

\section{Concluding Remarks}

In the present study, the accuracy of the Cowling approximation
has been investigated for rapidly and differentially 
rotating stars in Newtonian gravity. We have adopted a polytropic equation of 
state with polytropic indices $N=0.5$, 1.0, and 1.5
as the stellar models and estimated the accuracy of the approximation for the 
$m=2$ f- and r-mode oscillations. 

For the f-mode in rigidly rotating stars,
the tendency that the accuracy gets better for 
rapidly rotating stars has been reproduced.
On the other hand, when the angular momentum distribution of the star
is not uniform, the accuracy of the Cowling approximation becomes worse
when the rotation rate gets higher.
The accuracy of the approximation also depends on the equation of state.
For softer equations of state, the accuracy of the approximation becomes
relatively better even if the effect of differential rotation exists,
because of the central condensation of the matter.
Contrary, for stiffer equations of state,
which are usually used to describe the neutron stellar matter,
the accuracy of the approximation becomes
worse for rapidly and differentially rotating stars.
The reasons why the accuracy becomes worse can be considered as follows.
First, when the angular momentum distribution is not uniform,
the equilibrium configurations are highly deformed and the high density
region extends to the outer part of the equatorial region.
Second, when the angular momentum distribution is 
not uniform, the maximum amplitude of the eigenfunction appears well
inside the star where the equilibrium density is rather high.
It is possible to quantitatively confirm that the oscillations
become large in the high density region, by using a new quantity, $W_{\phi}$,
which is the integration of the density weighted by the eigenfunctions of
the oscillations.

As discussed in Introduction, it is extremely difficult to solve
oscillations of general relativistic rotating compact stars.
This is mainly because it is difficult to treat the perturbations of the 
metric, i.e. the perturbed gravitational potentials. Thus it is desirable 
to apply the Cowling approximation to general relativistic configurations 
and obtain reasonably accurate approximated solutions. 

Can we say anything about the applicability of the Cowling approximation to 
general relativistic configurations ?  It is widely believed that the accuracy 
of the Cowling approximation is better for general relativistic
configurations, because the strong gravity makes the effective central
condensation of the stellar matter higher.
This effect of gravitation is similar to the effect of the
increasing of the compressibility. As shown in the present paper, the
ascent of the compressibility achieves good accuracy of the approximation.
Therefore, in GR, it is expected that the accuracy of the 
Cowling approximation would become better than that for Newtonian gravity.

%
%

However, also in the case of GR, the same problem
accompanied with differential rotation will happen.
When we apply the Cowling approximation in differentially rotating
relativistic stars, the accuracy of the approximation will depend on the
positive effect of strong gravity and the negative effect of differential
rotation.
The detailed study must be carried out to discuss about the accuracy of
such an approximation in differentially rotating compact stars.
As mentioned in the last section, however, the Cowling approximation
works very well in the calculations of r-mode oscillations
(also see, Yoshida \& Lee 2002).
Hence, it is good idea to apply the approximation to obtain the
eigenfrequencies of r-modes in general relativity.
Also it is shown that the accuracy of
the Cowling approximation becomes better for higher modes (Jackson 1995),
therefore to compute those higher modes by using this approximation is
one of the realistic usages of the Cowling approximation.

\section*{Acknowledgments}

The present author thanks Yoshiharu Eriguchi for useful discussion,
comments, and continuous encouragements. He would like to thank Keisuke
Taniguchi and Koji Uryu for their careful reading and useful comments.
He thanks Yasufumi Kojima for his comments.
This work was in part supported by JSPS Research Fellowship for Young
Scientists.


\label{lastpage}

\end{document}